\documentclass[12pt,preprint,url]{aastex}
\usepackage{emulateapj5}
\usepackage{amsbsy}
\usepackage{amsmath}
\newcommand{\be}{\begin{equation}}
\newcommand{\ee}{\end{equation}}

\newcommand{\vn}{{\bf v}_{n}}
\newcommand{\vns}{{\bf v}_{ns}}
\newcommand{\vs}{{\bf v}_{s}}

\newcommand{\om}{\boldsymbol{\omega}}

\newcommand{\Rey}{{\it Re}}

\def\ltsima{$\; \buildrel < \over \sim \;$}

\def\lsim{\lower.5ex\hbox{\ltsima}}
\def\gtsima{$\; \buildrel > \over \sim \;$}
\def\gsim{\lower.5ex\hbox{\gtsima}}

\begin{document}
\title{Superfluid turbulence and pulsar glitch statistics}

\author{A. Melatos\altaffilmark{1} and C. Peralta\altaffilmark{1,}\altaffilmark{2} }

\email{a.melatos@physics.unimelb.edu.au}

\altaffiltext{1}{School of Physics, University of Melbourne,
Parkville, VIC 3010, Australia}

\altaffiltext{2}{Departamento de F\'{\i}sica, Escuela de Ciencias,
Universidad de Oriente, Cuman\'a, Venezuela}



\begin{abstract}
\noindent 
Experimental evidence is reviewed for the existence of
superfluid turbulence in a differentially rotating, spherical shell
at high Reynolds numbers ($\Rey\gsim 10^3$), such as the outer core of 
a neutron star.
It is shown that torque variability increases with $\Rey$, suggesting
that glitch activity in radio pulsars may be a function of 
$\Rey$ as well. The $\Rey$ distribution
of the $67$ glitching radio pulsars with characteristic ages
$\tau_c \leq 10^6$ {\rm yr} is constructed from radio timing 
data and cooling curves and compared with the $\Rey$ distribution of
all $348$ known pulsars with $\tau_c \leq 10^6$ {\rm yr}. The two
distributions are different, with a Kolmogorov-Smirnov probability
$\geq 1 - 3.9 \times 10^{-3}$. The conclusion holds for
(modified) Urca and nonstandard cooling, and for
Newtonian and superfluid viscosities.
\end{abstract}


\section{Introduction}
The known number of glitching radio pulsars has quadrupled in
the last four years, as has the total number of glitches,
in the wake of the Parkes Multibeam Survey 
and improved multifrequency timing solutions \citep{peralta06}.
Out of a total population of $1730$ pulsars,
$101$ have experienced a total of $287$ glitches
\citep{wan00,Kra03,catalog04,js06}. The enlarged
data set invites an updated statistical
study. Previous analyses identified
certain empirical trends, e.g., glitch activity
and healing
parameter versus characteristic age \citep{sl96,uo99,wan00}, and
glitch amplitude versus recurrence time \citep{mmwgz06}, while
refuting certain theoretically predicted correlations, 
e.g., glitch amplitude versus
spin period $P$, period derivative $\dot{P}$,
and total spin down since the last glitch (``reservoir model") \citep{sl96,wan00}.

On the theoretical front, 
the global pattern of superfluid circulation
in the outer core of a neutron star was recently computed
numerically for the first time \citep{pmgo05a,pmgo06a,pmgo06b}.
In general, a fluid
contained in a differentially rotating, spherical shell flows
unsteadily under a wide range of conditions, especially in thick
shells and at high Reynolds numbers $\Rey \gsim 10^3$ \citep{je00}. 
Such unsteady
flow is observed in numerical and laboratory experiments with
viscous fluids, e.g., transitions
between Taylor vortex states \citep{mt87b},
Taylor vortex oscillations \citep{h98}, relaminarization \citep{nst05},
and nonaxisymmetric modes on the route to high-$\Rey$ turbulence,
like Taylor-G\"ortler vortices \citep{li04}.
Time-dependent flow is also observed in superfluids
like helium II, e.g., torque ``steps"
and quasiperiodic
torque oscillations in Couette and spin-up experiments \citep{tsatsa80}. 

In this Letter, we combine theory and
observation to test whether superfluid turbulence
exists in neutron stars, as originally proposed by
\citet{g70}. In \S\ref{sec:hydro},
we present numerical simulations of a
$^1$S$_0$-paired neutron superfluid in the outer core of a 
neutron star as the crust spins down
electromagnetically. We show that the torque
is steady for $\Rey \lsim 10^3$ and unsteady 
(e.g., oscillatory) otherwise.
In \S\ref{sec:reynolds}, we combine
standard cooling curves with measurements of the characteristic
ages of radio pulsars to compute the temperature, viscosity, 
and hence $\Rey$ of the star. We show that
the $\Rey$ distributions of those pulsars that glitch, and
those that do not, are different --- a rare instance of regularity 
in the glitch phenomenon. The implications
for the existence of superfluid turbulence
in neutron stars are weighed critically in \S\ref{sec:discussion}.

\section{Hydrodynamic superfluid turbulence}
\label{sec:hydro}
Large-scale, time-dependent, meridional circulation, driven by Ekman
pumping, is a generic feature of viscous and superfluid flow
in spherical Couette geometry (i.e., a differentially rotating,
spherical shell); see \citet{je00} for a review. The complexity
of the global circulation pattern increases with $\Rey$. When $\Rey$
is large enough, the flow becomes nonaxisymmetric and ultimately
chaotic, e.g., \citet{ntz02a}.

\begin{figure*}
\epsscale{0.7}
\plotone{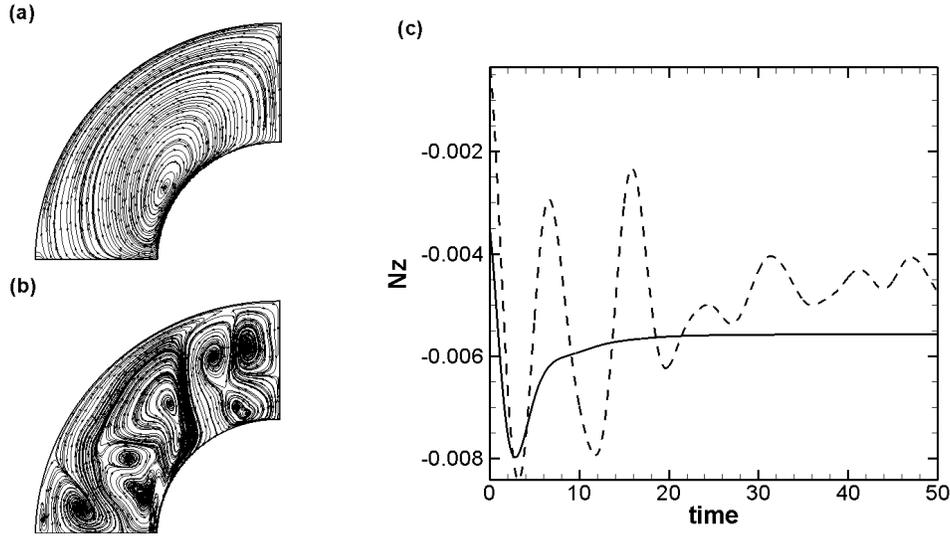}
\caption{Snapshot at $t=50 \Omega^{-1}$ of meridional streamlines 
(computed by integrating the in-plane velocity vector)
for the normal fluid component
in spherical Couette flow in the outer core of a neutron star
for (a) $\Rey=3 \times 10^2$ and (b)  $\Rey=3 \times 10^4$.
(c) Evolution of the outer torque on the crust (in units of $\rho R^5 \Omega^2$)
for $\Rey=3 \times 10^2$ (solid curve) and $\Rey=3 \times 10^4$ (dashed curve).
The simulation parameters are mutually ordered as in the 
outer core of a neutron star but exaggerated for computational
tractability, viz., crust-core angular velocity shear
$\Delta \Omega / \Omega = 0.3$, 
dimensionless shell thickness
$\delta = 0.5$, stiffness parameter $\nu_s = 10^{-5}  R^2 \Omega \ll \nu_n$, Gorter-Mellink constant
$A^\prime = 5.8 \times 10^{-2}$, and superfluid density fraction
$\rho_s / \rho = 0.99$. We follow exactly the notation and definitions in \citet{pmgo05a}.}
\label{fig:re11}
\end{figure*}

Figure \ref{fig:re11} compares spherical Couette flow of a two-component,
Hall-Vinen-Bekarevich-Khalatnikov (HVBK) superfluid at low
and high $\Rey$. Meridional streamlines of the normal
(i.e., torque generating) component are plotted for $\Rey = 3 \times 10^2$
and $3 \times 10^4$ in Figures 1a and 1b respectively.
The flow is computed by integrating the HVBK equations
\citep{hr77} numerically using a pseudospectral collocation
method \citep{canuto88} with spectral filtering 
to suppress high spatial frequencies,
especially at high $\Rey$ \citep{pmgo05a,peralta06}.
We adopt no-slip and no-penetration boundary conditions
for the normal component (velocity $\vn$), perfect slip for 
the superfluid component (i.e., no pinning, for
numerical stability; velocity $\vs$), and a Stokes flow 
(with $\vn = \vs$) initially. The mutual
friction force is of the isotropic, 
Gorter-Mellink form ($\propto v_{ns}^2 \vns$,  with
$\vns=\vn-\vs$) \citep{gm49}, because
the meridional counterflow in Figures \ref{fig:re11}a and \ref{fig:re11}b
excites Kelvin waves via
the Donnelly-Glaberson instability, converting 
the rectilinear vortex
array into a vortex tangle \citep{sbd83,tab03,pmgo06b}.
The superfluid component also feels a vortex tension force 
$\propto \mbox{\boldmath$\omega$}_{s} \times (\nabla \times \hat{\mbox{\boldmath$\omega$}}_{s})$, with $\mbox{\boldmath$\omega$}_{s} = \nabla \times \vs$.

The meridional circulation pattern in Figure \ref{fig:re11}b
is more complicated than in Figure \ref{fig:re11}a; seven
cells are visible for $\Rey = 3 \times 10^4$, compared to only one
cell for $\Rey = 3 \times 10^2$.
The azimuthally averaged differential torque
$dN_z / d(\cos \theta)$ \citep{pmgo06b}
is a monotonic function of latitude
for $\Rey = 3 \times 10^2$
but peaks near the (moving) secondary vortices (at $\theta \approx 30$\ensuremath{^\circ})
in Figure \ref{fig:re11}b for $\Rey = 3 \times 10^4$.

The torque is variable at high $\Rey$, exhibiting
persistent oscillations under steady differential rotation, 
whose amplitude depends on $\Rey$ \citep{pmgo05a}. Figure \ref{fig:re11}c
displays the torque on the outer sphere (stellar crust)
as a function of time. For $t \geq 20 \, \Omega^{-1}$, with
$\Omega=2\pi/P$, after
initial transients die away, we see that the torque is asymptotically
steady for $\Rey = 3 \times 10^2$ (solid curve) 
and oscillatory for $\Rey = 3 \times 10^4$ (dashed curve).
The variable torque leads naturally to rotational irregularities. 
The variability time-scale ($\sim \Omega^{-1}$) is too
fast to account for pulsar timing noise directly. 
However, the
torque oscillates quasiperiodically, not periodically, 
so $\Omega$ wanders stochastically when integrated over long times 
[c.f. microjumps
and oscillations superimposed on a random walk;
\citet{bghnpw72,hthesis02}].

Of course, we wish to investigate the flow at 
$\Rey \gsim 10^9$, the realistic $\Rey$ regime for radio
pulsars \citep{mm05}. Unfortunately, our simulations cannot
handle this. Even for $\Rey = 10^5$, and
with heavy spectral filtering,
the flow is unresolved numerically;
the spectral coefficients do not decline monotonically
with polynomial order as they should \citep{peralta06}.
From laboratory experiments, however, it is clear 
what to expect qualitatively
as $\Rey$ increases: the flow passes through a sequence of 
bifurcations to nonaxisymmetric states like herringbone waves
and Stuart vortices at $\Rey \sim 10^6$ \citep{je00}, before
developing
fully into scale-free, Kolmogorov-like turbulence at $\Rey \gsim 10^7$,
which itself is not always isotropic at small scales \citep{dcb91}.
Turbulent flow in He II
at $\Rey \gsim 10^5$ often contains a vortex tangle \citep{bsbd97}.
Moreover, in experiments with He II, eddies in the normal and
superfluid components match at scales larger than
the average vortex separation \citep{bsbd97},
implying that, macroscopically, high-$\Rey$ superfluid turbulence
behaves similarly to high-$\Rey$ classical turbulence.
Hence, in neutron stars ($\Rey \gsim 10^9$), the
vortex tangle in the outer core may match its
vorticity to the normal component.
The vortex tension modifies this picture
somewhat by contributing ``stiffness" in regions where 
$\om_s$ builds up (e.g., at the smallest
scales in the Kolmogorov cascade).
The largest eddies
in fully developed turbulence, where the kinetic energy mainly resides,
are nonaxisymmetric and move ``jerkily",
so the net torque fluctuates substantially.

\section{$\Rey$ distribution of glitching pulsars}
\label{sec:reynolds}
Given that the Reynolds number is a key factor governing the
variability of the global superfluid flow
in a neutron star, and hence
the star's rotation, it is interesting to test
whether the amplitude and rate of incidence
of rotational irregularities depends
on it. From the definition
\begin{equation}
\label{eq:rey1} \Rey = {R^2 \Omega}/{\nu_n},
\end{equation}
where $R$ is the stellar radius and $\nu_n$ is the kinematic
viscosity of the normal component,
it is clear that $\Rey$ can be measured in principle, if we know how
$\nu_n$ depends on the density $\rho$ and temperature $T$
in the outer core. Using the Newtonian
viscosity formula derived by \citet{cl87}, due 
to neutron-neutron scattering, we find
\begin{equation}
\label{eq:rey2}
\Rey = 1.8 \times 10^{11} 
( {T}/{10^8 \, {\rm K}})^2
( {\Omega}/{10^2 \, {\rm rad} \, {\rm s}^{-1}}),
\end{equation}
with $\rho = 2.8 \times 10^{12}$ {\rm g} {\rm cm}$^{-3}$
(normal component)
and $R = 10^6$ {\rm cm}. For electron-electron scattering 
in a superfluid, the multiplicative factor
is $6.0 \times 10^{10}$ \citep{acg05}.
In writing equations (\ref{eq:rey1}) and (\ref{eq:rey2}),
we imagine a model where the outer core extends
over the density range ($2$--$4$) $\times 10^{14}$ {\rm g} {\rm cm}$^{-3}$,
where the superfluid behaves hydrodynamically (i.e., weak
bulk pinning) and isotropically ($^1$S$_0$-paired). We assume that the 
normal component constitutes $1$ \% of the total density,
to account for condensate depletion
by nonideal effects. We do not treat stratification in the 
simulations to keep them tractable, although stratification 
is known to be important in suppressing meridional 
circulation \citep{ae96,pmgo06b}. 

The core temperature $T$ is related to the surface
temperature $T_s$, e.g., via the two-zone, heat-blanket
model of \citet{gpe82}, which gives 
$(T/10^8 \, {\rm K}) = 1.29 ({T_s}/{10^6 \, {\rm K}})^{1.8}$.
Surface temperatures have been
measured for $11$ neutron stars, by fitting their thermal X-ray
emission to hydrogen or heavy-element atmospheres \citep{ll02,pp04}.
However, the sample is too small to analyse statistically;
only three of these objects exhibit glitches,
for example. Instead, we are forced to estimate $T_s$ from
the characteristic age $\tau_c = P/(2\dot{P})$, 
combined with theoretical
cooling curves, calibrated against the $11$ objects above 
\citep{page98}. We consider three cooling mechanisms:
(I) standard neutrino cooling, via the Urca
and modified Urca processes \citep{hgrr70}, (II) nonstandard
(fast) cooling due to paired neutron superfluidity \citep{ao85a},
and (III) nonstandard cooling with the superfluid energy gap
fitted empirically to observations of thermal emission
\citep{gys94,lp96}. 

Figure \ref{fig:re13}a displays the raw $\Rey$ distributions of the
$67$ glitching
pulsars with $\tau_c \leq 10^6$ {\rm yr} (green histogram) and
all $348$  known pulsars with
$\tau_c \leq 10^6$ {\rm yr} (red histogram), including 
the $67$ glitchers. The objects are sorted
into ten bins of equal logarithmic width $0.37$ dex in the interval
$9.4 \leq \log_{10} (\Rey) \leq 13.1$. Standard cooling (I) 
and Newtonian viscosity formulas are applied to both samples.
We restrict attention to objects 
with $\tau_c \leq 10^6$ {\rm yr}, because
the cooling curves compiled by \citet{page98} drop
away precipitiously above this age, as neutrino cooling
gives way to photon cooling (which depends sensitively on
the uncertain composition of the stellar atmosphere).

The histogram in Figure \ref{fig:re13}a is redrawn for 
nonstandard cooling (II and III) in Figures \ref{fig:re13}b
and \ref{fig:re13}c, with bin widths of $0.30$ dex
and $0.96$ dex, respectively. Identical (after rescaling $\Rey$)
distributions are obtained if the superfluid electron-electron 
scattering viscosity is used instead.

The $\Rey$ distribution of the glitching pulsars differs clearly
from that of the total population. One sees immediately,
from Figure \ref{fig:re13}a,
that most of the glitching pulsars have $10^{11} \lsim \Rey \lsim 10^{12}$,
whereas the pulsar population peaks at $10^{10} \lsim \Rey \lsim 10^{11}$.
To quantify this, we perform a Kolmogorov-Smirnov (K-S) test on
the cumulative distributions constructed from Figure \ref{fig:re13}a
(to circumvent binning bias). The K-S statistic
is $ D = 0.3$, yielding a probability $p=1-1.6 \times 10^{-5}$
that the two populations are drawn from different distributions.
Similar conclusions follow for nonstandard cooling of types II and III,
and for superfluid electron-electron scattering viscosity
($p$ does not change when $\Rey$ is rescaled multiplicatively).
The associated K-S-probabilities are quoted in Table \ref{table:t1}.

Is Figure \ref{fig:re13} just a restatement of the well known
trend that glitching pulsars tend to be young? Partly, but
not entirely. The period distribution of glitching pulsars differs
clearly from that of the total population, with a K-S probability
$p=0.994$. The maximum separation between the cumulative
$P$ distributions occurs at an intermediate period
$P=0.32$ {\rm s}, ruling out a pure age effect.
[Observational studies of kick velocities and pulsars in supernova remnants
suggest that the distribution of the birth periods
is quite flat in the range $0.05-0.5$ {\rm s}; see
\citet{fk06}, \citet{nr07}, and references therein.] 
It appears that $P$ and $\tau_c$ 
control glitch
behaviour independently (perhaps, but not necessarily, through
$\Rey$). Interestingly, glitching pulsars are distributed
differently in $\tau_c$ and $\Rey$, peaking
at $\tau_c \sim 10^{6}$ {\rm yr} (oldest objects)
and $\Rey \sim 10^{11}$ (intermediate objects) respectively.

Given that $\Rey$ distributions of glitching
and nonglitching pulsars are different, it is natural 
to ask whether the rate and amplitude
of glitch activity are simple functions of $\Rey$. 
There are
several possible measures of glitch activity in an individual pulsar.
We pick three: (i) the activity parameter 
$A_g = (N_g \Delta \Omega_g)/(t_g \Omega)$, where $N_g$ is the number of glitches observed over
a total observation time $t_g$ (taken to be
the time since discovery, in the absence of complete data),
and $\Delta \Omega_g$ is
the accumulated change in $\Omega$ due to glitches \citep{ml90};
(ii) $\Delta \Omega_g/\Omega$; and (iii) the {\it mean} recurrence time
$t_r$ between glitches, normalized by $\tau_c$
(in $11$ out of $67$ objects, this is an
upper limit, as they have glitched once only). 
Interestingly, none of these measures show a clear
trend with $\Rey$, leading to scatter plots. This suggests that $\Rey$ controls
the threshold for glitch activity, rather than its rate
and amplitude, which is qualitatively consistent with \S\ref{sec:hydro}. 
For example,
one can imagine a $\Rey$ threshold (when
turbulence sets in), which must be exceeded for glitches to
occur at all, and a separate ``rate" threshold 
(perhaps controlled by the Rossby number, proportional
to the crust-core shear),
which determines how often the pulsar glitches
per unit observation time (and hence $A_g$).
Likewise, $t_r/\tau_c$ and $\Delta \Omega_g /\Omega$ are related physicallly
to the rate at which crust-core  shear
builds up, which is independent of $\Rey$.
Also, $A_g$ is not dimensionless,
so it is not surprising that factors other than $\Rey$ affect it.
Note that $\Rey$ decreases with $\tau_c$ but varies
independently with $\Omega$, so the $A_g$-$\tau_c$ correlation
\citep{wan00} is washed out in an $A_g$-$\Rey$ plot.

\section{Discussion}
\label{sec:discussion}
A superfluid (or, indeed, a Newtonian fluid) contained
in a differentially rotating, spherical shell circulates meridionally
and exerts a time-dependent viscous torque on the outer shell. The
vigour of the circulation, and the variability of the torque,
increase with $\Rey$;  the flow is
steady for $\Rey \lsim 10^2$ and chaotic for $\Rey \gsim 10^5$.
Spherical Couette flow is an idealised model of the superfluid
outer core of a neutron star, in which the outer sphere is the
stellar crust, which spins down electromagnetically. It is 
therefore interesting to test whether the incidence
of pulsar rotational irregularities like
glitches depends on $\Rey$. We show in
\S\ref{sec:reynolds} that this is indeed the case: the
distribution of glitching and nonglitching pulsars 
with $\tau_c \leq 10^6$ {\rm yr} is different, with K-S probability
$\geq 1 - 3.9 \times 10^{-3}$.
This new fact joins the $A_g$-versus-$\tau_c$ correlation
as one the few observed regularities in glitch phenomenology.

\begin{figure*}
\begin{center}
\epsscale{1.02}
\plotone{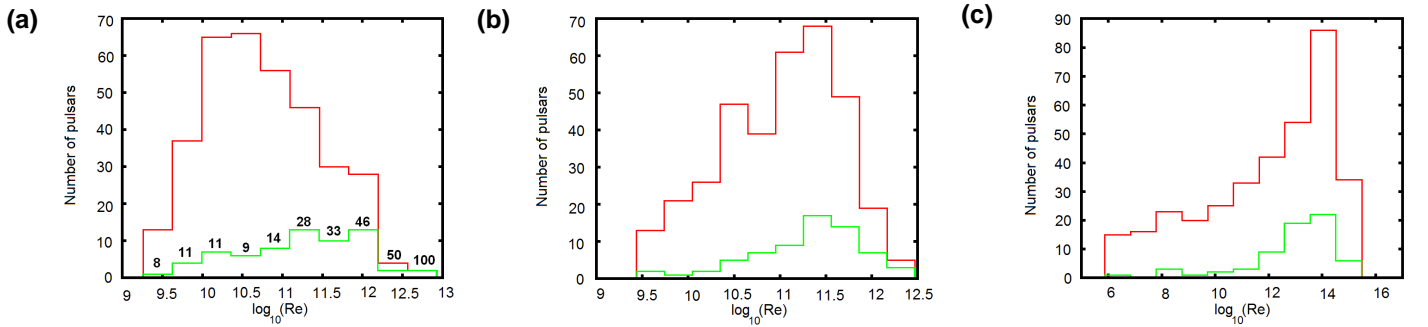}
\caption{Reynolds number distribution for the $67$ glitchers (green curve) and all $348$ pulsars
(red curve, including the $67$ glitchers) with characteristic age $\tau_c \leq 10^6$ {\rm yr},
using (a) standard neutrino cooling, (b) nonstandard
(fast) cooling due to paired neutron superfluidity,
and (c) nonstandard cooling with the superfluid energy gap
fitted empirically to observations of thermal emission.
The data are divided into $10$ bins of equal logarithmic width (a) $0.37$,
(b) $0.30$, and (c) $0.96$. The percentage of objects that
glitch in each bin, for standard neutrino cooling, 
is recorded above the histogram in (a).}
\label{fig:re13}
\end{center}
\end{figure*}

An immediate worry is that the distributions in Figure \ref{fig:re13}
are contaminated by some observational selection effect.
Our estimates of $\Rey$ are uncertain in two ways. First, $\tau_c$
is the characteristic spin-down age of the pulsar, not its true
age. For example, five young pulsars with $\tau_c \leq 1.7 \times 10^3$ {\rm yr}
have braking index $n < 3$, whereas the formula $\tau_c = P/(2\dot{P})$ assumes $n=3$
\citep{m97,lkgk06}. Second, the theoretical cooling curves depend
sensitively on the superfluid energy gap in the outer core, which
is poorly known (although admittedly calibrated
by $11$ pulsars with observed thermal emission). However,
the above uncertainties
do {\it not} affect the conclusion that the $\Rey$ distributions
of the glitching and nonglitching populations in Figure \ref{fig:re13}
are different, because $T$ and hence $\Rey$
are calculated in the same way for both samples. Indeed, it
is hard to imagine that our ability to detect glitches in
radio timing data is a function of $\Rey$.

Physically, the conclusions from Figure \ref{fig:re13}
are at once natural yet surprising. It is well known that torque
variability increases with $\Rey$ in a differentially rotating
superfluid in laboratory and numerical experiments
(\S\ref{sec:hydro}). However,
it is strange that pulsars behave differently
at $\Rey \lsim 10^{10}$ (few glitches) and $\Rey \gsim 10^{11}$
(many glitches). Naively, Kolmogorov turbulence should
be scale-free, fully developed, and {\it indistinguishable} at
these Reynolds numbers.
There are at least two ways to explain this. 
First, processes involving
microscopic superfluid turbulence, e.g., pinning \citep{apas84}
and vortex tangle formation \citep{pmgo06b}, may
depend sensitively on $\Rey$ in the range $10^{10} \leq \Rey \leq 10^{11}$.
Second, it is possible that theoretical estimates of $\Rey$
\citep{acg05,mm05} are $\sim 10^6$ times too high. If so, this reduces 
the peak of the 
glitching pulsar distribution to $\Rey \lsim 10^5$, exactly
where spherical Couette flow breaks up into
nonaxisymmetric modes before becoming chaotic (\S\ref{sec:hydro}).
We have no reason to doubt the $\nu_n$ carefully derived
in the literature. However, if the flow is turbulent, Reynolds
stresses may completely overwhelm viscous stresses, such that $\nu_n$ is replaced by
$\nu_n+A$, where $A$ satisfies $A \partial v_i/\partial x_j \approx
\langle \delta v_i \delta v_j \rangle$, $v_i$ is the mean
velocity, and $\delta v_i$ is the fluctuating velocity
(zero average) \citep{pedlosky_book}. Typically, we have
$A/\nu_n \sim (\delta v/v)^2 \Rey \gg 1$. Turbulent Reynolds
stresses drastically reduce $\Rey$ in a range of 
geophysical and astrophysical applications \citep{pedlosky_book}.

Plainly, there exist several high-$\Rey$ pulsars that
do not glitch. Observational selection effects aside, 
there exist several theoretical 
mechanisms that may be responsible.
First, for spherical Couette flow
in the range $10^3 < \Rey < 10^7$, there are sizable
intervals of $\Rey$ where the flow is almost laminar,
interposed between intervals where it is turbulent \citep{nt05}.  
Second, for  $\Rey > 10^7$, where the turbulence is fully developed 
(Kolmogorov) and ``isotropic" when averaged over a turn-over time, the torque 
on the star is steadier than for organised modes (e.g., Taylor-G\"ortler vortices) 
near the onset of turbulence ($\Rey \sim 10^5$), which are
not isotropic on average. Third,
if atmospheres and planetary interiors are any guide, the
turbulent Reynolds stresses depend sensitively on what
nonlinear modes are excited in the turbulence,
and the effective $\Rey$ may vary greatly (and unpredictably) 
from object to object, in ways not captured by (\ref{eq:rey2}).
Incidentally, none of the 
pulsars in our sample have low $\Rey$, even after
adjusting for turbulent Reynolds stresses; the minimum
is $\Rey \sim 10^3$. Hence there are no 
examples of low-$\Rey$ objects that glitch
unexpectedly.

In closing, we emphasize that the results of this paper do
{\it not} prove that superfluid turbulence exists in
a neutron star, let alone that it controls glitch behaviour.
However, considerable effort has been expended to identify
patterns in glitch behaviour, with limited success. The undeniable
empirical fact that the $\Rey$ distributions in Figure \ref{fig:re13} are
different is therefore intriguing, especially because $\Rey$ is 
a {\it dimensionless} quantity with {\it deep hydrodynamical significance},
and because the K-S comparison is robust
with respect to how $T$, $\nu_n$, and hence $\Rey$ are estimated
observationally. One hopes it will offer an enlightening clue
to glitch physics, whether or not superfluid turbulence
turns out to play a central role.
We also hope that the results presented here
will stimulate the ongoing observational campaign to measure
pulsar temperatures.

\acknowledgments 
This research was supported by
a postgraduate scholarship from the University of Melbourne
and the Albert Shimmins writing-up award.
It makes use of the ATNF pulsar catalogue
\url{http://www.atnf.csiro.au/research/pulsar/psrcat}
\citep{atnf_catalog_05} and unpublished glitch data
provided kindly by D. Lewis, M. Kramer, and
A. Lyne (private communications, 2005).
We thank the anonymous referee for helpful suggestions.


\begin{thebibliography}{49}
\expandafter\ifx\csname natexlab\endcsname\relax\def\natexlab#1{#1}\fi

\bibitem[{{Abney} \& {Epstein}(1996)}]{ae96}
{Abney}, M., \& {Epstein}, R.~I. 1996, J. Fluid Mech., 312, 327

\bibitem[{{Alpar} {et~al.}(1984){Alpar}, {Pines}, {Anderson}, \&
  {Shaham}}]{apas84}
{Alpar}, M.~A., {Pines}, D., {Anderson}, P.~W., \& {Shaham}, J. 1984, \apj,
  276, 325

\bibitem[{{Amundsen} \& {{\O}stgaard}(1985)}]{ao85a}
{Amundsen}, L., \& {{\O}stgaard}, E. 1985, Nucl. Phys. A, 437, 487

\bibitem[{{Andersson} {et~al.}(2005){Andersson}, {Comer}, \&
  {Glampedakis}}]{acg05}
{Andersson}, N., {Comer}, G.~L., \& {Glampedakis}, K. 2005, Nucl. Phys. A, 763,
  212

\bibitem[{{Barenghi} {et~al.}(1997){Barenghi}, {Samuels}, {Bauer}, \&
  {Donnelly}}]{bsbd97}
{Barenghi}, C.~F., {Samuels}, D.~C., {Bauer}, G.~H., \& {Donnelly}, R.~J. 1997,
  Phys. Fluids, 9, 2631

\bibitem[{{Boynton} {et~al.}(1972){Boynton}, {Groth}, {Hutchinson}, {Nanos},
  {Partridge}, \& {Wilkinson}}]{bghnpw72}
{Boynton}, P.~E., {Groth}, E.~J., {Hutchinson}, D.~P., {Nanos}, G.~P.,
  {Partridge}, R.~B., \& {Wilkinson}, D.~T. 1972, \apj, 175, 217

\bibitem[{{Canuto} {et~al.}(1988){Canuto}, {Hussaini}, {Quarteroni}, \&
  {Zang}}]{canuto88}
{Canuto}, C., {Hussaini}, M., {Quarteroni}, A., \& {Zang}, T. 1988, {Spectral
  Methods in Fluid Dynamics} (Springer-Verlag)

\bibitem[{{Cutler} \& {Lindblom}(1987)}]{cl87}
{Cutler}, C., \& {Lindblom}, L. 1987, \apj, 314, 234

\bibitem[{{Douady} {et~al.}(1991){Douady}, {Couder}, \& {Brachet}}]{dcb91}
{Douady}, S., {Couder}, Y., \& {Brachet}, M.~E. 1991, \prl, 67, 983

\bibitem[{{Faucher-Gigu{\`e}re} \& {Kaspi}(2006)}]{fk06}
{Faucher-Gigu{\`e}re}, C.-A., \& {Kaspi}, V.~M. 2006, \apj, 643, 332

\bibitem[{{Gnedin} {et~al.}(1994){Gnedin}, {Yakovlev}, \& {Shibanov}}]{gys94}
{Gnedin}, O.~Y., {Yakovlev}, D.~G., \& {Shibanov}, Y.~A. 1994, Astron. Lett.,
  20, 409

\bibitem[{{Gorter} \& {Mellink}(1949)}]{gm49}
{Gorter}, C.~J., \& {Mellink}, J.~H. 1949, Physica, 85, 285

\bibitem[{{Greenstein}(1970)}]{g70}
{Greenstein}, G. 1970, \nat, 227, 791

\bibitem[{{Gudmundsson} {et~al.}(1982){Gudmundsson}, {Pethick}, \&
  {Epstein}}]{gpe82}
{Gudmundsson}, E.~H., {Pethick}, C.~J., \& {Epstein}, R.~I. 1982, \apjl, 259,
  L19

\bibitem[{{Hills} \& {Roberts}(1977)}]{hr77}
{Hills}, R.~N., \& {Roberts}, P.~H. 1977, Arch. Rat. Mech. Anal., 66, 43

\bibitem[{{Hobbs}(2002)}]{hthesis02}
{Hobbs}, G. 2002, PhD thesis, University of Manchester

\bibitem[{{Hobbs} {et~al.}(2004){Hobbs}, {Manchester}, {Teoh}, \&
  {Hobbs}}]{catalog04}
{Hobbs}, G., {Manchester}, R., {Teoh}, A., \& {Hobbs}, M. 2004, in IAU
  Symposium, 139--+

\bibitem[{{Hoffberg} {et~al.}(1970){Hoffberg}, {Glassgold}, {Richardson}, \&
  {Ruderman}}]{hgrr70}
{Hoffberg}, M., {Glassgold}, A.~E., {Richardson}, R.~W., \& {Ruderman}, M.
  1970, \prl, 24, 775

\bibitem[{{Hollerbach}(1998)}]{h98}
{Hollerbach}, R. 1998, \prl, 81, 3132

\bibitem[{{Janssen} \& {Stappers}(2006)}]{js06}
{Janssen}, G.~H., \& {Stappers}, B.~W. 2006, astro-ph/0607260

\bibitem[{{Junk} \& {Egbers}(2000)}]{je00}
{Junk}, M., \& {Egbers}, C. 2000, LNP Vol.~549: Physics of Rotating Fluids,
  549, 215

\bibitem[{{Krawczyk} {et~al.}(2003){Krawczyk}, {Lyne}, {Gil}, \&
  {Joshi}}]{Kra03}
{Krawczyk}, A., {Lyne}, A.~G., {Gil}, J.~A., \& {Joshi}, B.~C. 2003, \mnras,
  340, 1087

\bibitem[{{Larson} \& {Link}(2002)}]{ll02}
{Larson}, M., \& {Link}, B. 2002, \mnras, 333, 613

\bibitem[{{Levenfish} \& {Yakovlev}(1996)}]{lp96}
{Levenfish}, K.~P., \& {Yakovlev}, D.~G. 1996, Astron. Lett., 22, 49

\bibitem[{{Li}(2004)}]{li04}
{Li}, Y. 2004, Sci. China Ser. A, 47, 81

\bibitem[{{Livingstone} {et~al.}(2006){Livingstone}, {Kaspi}, {Gotthelf}, \&
  {Kuiper}}]{lkgk06}
{Livingstone}, M.~A., {Kaspi}, V.~M., {Gotthelf}, E.~V., \& {Kuiper}, L. 2006,
  \apj, 647, 1286

\bibitem[{{Manchester} {et~al.}(2005){Manchester}, {Hobbs}, {Teoh}, \&
  {Hobbs}}]{atnf_catalog_05}
{Manchester}, R.~N., {Hobbs}, G.~B., {Teoh}, A., \& {Hobbs}, M. 2005, Astron.
  J., 129, 1993

\bibitem[{{Marcus} \& {Tuckerman}(1987)}]{mt87b}
{Marcus}, P., \& {Tuckerman}, L. 1987, J. Fluid Mech., 185, 31

\bibitem[{{Mastrano} \& {Melatos}(2005)}]{mm05}
{Mastrano}, A., \& {Melatos}, A. 2005, \mnras, 361, 927

\bibitem[{{McKenna} \& {Lyne}(1990)}]{ml90}
{McKenna}, J., \& {Lyne}, A.~G. 1990, Nature, 343, 349

\bibitem[{{Melatos}(1997)}]{m97}
{Melatos}, A. 1997, \mnras, 288, 1049

\bibitem[{{Middleditch} {et~al.}(2006){Middleditch}, {Marshall}, {Wang},
  {Gotthelf}, \& {Zhang}}]{mmwgz06}
{Middleditch}, J., {Marshall}, F.~E., {Wang}, Q.~D., {Gotthelf}, E.~V., \&
  {Zhang}, W. 2006, astro-ph/0605007

\bibitem[{{Nakabayashi} {et~al.}(2005){Nakabayashi}, {Sha}, \&
  {Tsuchida}}]{nst05}
{Nakabayashi}, K., {Sha}, W., \& {Tsuchida}, Y. 2005, J. Fluid Mech., 534, 327

\bibitem[{{Nakabayashi} \& {Tsuchida}(2005)}]{nt05}
{Nakabayashi}, K., \& {Tsuchida}, Y. 2005, Phys. Fluids, 17, 4110

\bibitem[{{Nakabayashi} {et~al.}(2002){Nakabayashi}, {Tsuchida}, \&
  {Zheng}}]{ntz02a}
{Nakabayashi}, K., {Tsuchida}, Y., \& {Zheng}, Z. 2002, Phys. Fluids, 14, 3963

\bibitem[{{Ng} \& {Romani}(2007)}]{nr07}
{Ng}, C.~., \& {Romani}, R.~W. 2007, ArXiv Astrophysics e-prints

\bibitem[{{Page}(1998)}]{page98}
{Page}, D. 1998, in Neutron Stars and Pulsars: Thirty Years after the
  Discovery, 183--+

\bibitem[{{Page} {et~al.}(2004){Page}, {Lattimer}, {Prakash}, \&
  {Steiner}}]{pp04}
{Page}, D., {Lattimer}, J.~M., {Prakash}, M., \& {Steiner}, A.~W. 2004, \apjs,
  155, 623

\bibitem[{{Pedlosky}(1982)}]{pedlosky_book}
{Pedlosky}, J. 1982, {Geophysical fluid dynamics} (New York and Berlin,
  Springer-Verlag, 1982.~636 p.)

\bibitem[{{Peralta}(2006)}]{peralta06}
{Peralta}, C. 2006, PhD thesis, University of Melbourne

\bibitem[{{Peralta} {et~al.}(2005){Peralta}, {Melatos}, {Giacobello}, \&
  {Ooi}}]{pmgo05a}
{Peralta}, C., {Melatos}, A., {Giacobello}, M., \& {Ooi}, A. 2005, \apj, 635,
  1224

\bibitem[{{Peralta} {et~al.}(2006{\natexlab{a}}){Peralta}, {Melatos},
  {Giacobello}, \& {Ooi}}]{pmgo06a}
---. 2006{\natexlab{a}}, \apjl, 644, L53

\bibitem[{{Peralta} {et~al.}(2006{\natexlab{b}}){Peralta}, {Melatos},
  {Giacobello}, \& {Ooi}}]{pmgo06b}
---. 2006{\natexlab{b}}, \apj, 651, 1079

\bibitem[{{Shemar} \& {Lyne}(1996)}]{sl96}
{Shemar}, S.~L., \& {Lyne}, A.~G. 1996, \mnras, 282, 677

\bibitem[{{Swanson} {et~al.}(1983){Swanson}, {Barenghi}, \& {Donnelly}}]{sbd83}
{Swanson}, C.~E., {Barenghi}, C.~F., \& {Donnelly}, R.~J. 1983, \prl, 50, 190

\bibitem[{{Tsakadze} \& {Tsakadze}(1980)}]{tsatsa80}
{Tsakadze}, J.~S., \& {Tsakadze}, S.~J. 1980, J. Low Temp. Phys., 39, 649

\bibitem[{{Tsubota} {et~al.}(2003){Tsubota}, {Araki}, \& {Barenghi}}]{tab03}
{Tsubota}, M., {Araki}, T., \& {Barenghi}, C.~F. 2003, \prl, 90, 205301

\bibitem[{{Urama} \& {Okeke}(1999)}]{uo99}
{Urama}, J.~O., \& {Okeke}, P.~N. 1999, \mnras, 310, 313

\bibitem[{{Wang} {et~al.}(2000){Wang}, {Manchester}, {Pace}, {Bailes}, {Kaspi},
  {Stappers}, \& {Lyne}}]{wan00}
{Wang}, N., {Manchester}, R.~N., {Pace}, R.~T., {Bailes}, M., {Kaspi}, V.~M.,
  {Stappers}, B.~W., \& {Lyne}, A.~G. 2000, \mnras, 317, 843

\end{thebibliography}

\begin{table}
\begin{center}
\begin{tabular}{ccc}
\hline
\hline
 Cooling &  $p$ \\ \hline
 I &  $1-6.4 \times 10^{-5}$  \\ 
 II & $1 - 3.9 \times 10^{-3}$  \\
 III & $1 - 2.5 \times 10^{-3}$ \\ \hline
\end{tabular}
\end{center}
\caption{K-S probabilities $p$ for Newtonian viscosity
and three cooling mechanisms.}
\label{table:t1}
\end{table}

\end{document}